\documentclass[twocolumn]{aastex631}

\usepackage{graphicx}	
\usepackage{amsmath}	
\usepackage{nccmath}
\usepackage{amssymb}	
\usepackage{multirow}

\shorttitle{Periods of pulsars from synchro-curvature spectra}
\shortauthors{\'I\~niguez-Pascual, Torres, \& Vigan\`o }

\begin{document}

\title{Inferring pulsar periods from synchro-curvature spectra}

\author{Daniel \'I\~niguez-Pascual}\thanks{iniguez@ice.csic.es}
\affiliation{Institute of Space Sciences (ICE, CSIC), Campus UAB, Carrer de Can Magrans s/n, 08193 Barcelona, Spain}
\affiliation{Institut d’Estudis Espacials de Catalunya (IEEC), 08034 Barcelona, Spain}

\author{Diego F. Torres}\thanks{dtorres@ice.csic.es}
\affiliation{Institute of Space Sciences (ICE, CSIC), Campus UAB, Carrer de Can Magrans s/n, 08193 Barcelona, Spain}
\affiliation{Institut d’Estudis Espacials de Catalunya (IEEC), 08034 Barcelona, Spain}
\affiliation{Institució Catalana de Recerca i Estudis Avançats (ICREA), E-08010 Barcelona, Spain}

\author{Daniele Vigan\`o}\thanks{vigano@ice.csic.es}
\affiliation{Institute of Space Sciences (ICE, CSIC), Campus UAB, Carrer de Can Magrans s/n, 08193 Barcelona, Spain}
\affiliation{Institut d’Estudis Espacials de Catalunya (IEEC), 08034 Barcelona, Spain}

\begin{abstract}
The period and the period derivative of a pulsar are critical magnitudes for defining the properties of the magnetospheric size and plasma dynamics. The pulsar light cylinder, the magnetic field intensity nearby it, and the curvature radius all depend on these timing properties, and shape the observed high-energy synchro-curvature emission. Therefore, the radiative properties of pulsars are inextricably linked to them. This fact poses the question of how well does a given pulsar's spectral energy distribution embeds information of the timing parameters, and if so, whether we can deduce them if they have not been measured directly. This is relevant to possibly constrain the timing properties of potential pulsar candidates among unidentified $\gamma$-ray sources. 
We consider well-measured pulsar spectra blinding us from the knowledge of their timing properties, and address this question by using our radiative synchro-curvature model that was proven able to fit the observed spectra of the pulsar population. We find that in the majority of the cases studied (8/13), the spin period is constrained within a range of about one order of magnitude, within which the real period lies. In the other cases, there is degeneracy and no period range can be constrained. This can be used to facilitate the blind search of pulsed signals.
\end{abstract}


\section{Introduction}
\label{introduction}

Possible pulsar candidates among $\gamma$-ray unidentified sources can be signaled 
by comparing their spectral energy distributions (SEDs) with those typical of known pulsars (see, e.g., \cite{2fpc}).
This includes searching for slopes, curvature, cutoffs, or peak energies similar to those shown by members of the detected pulsar population.
Machine learning techniques have also been used to try to identify pulsar candidates. 
For instance,  \cite{SazParkinson16} used Random Forest and Logistic Regression to classify {\it Fermi}-LAT $\gamma$-ray sources into either pulsars or active galactic nuclei.

Complementary to the previous approaches, here we shall assume that a given SED is possibly pulsar-generated and explore whether we can actually determine what would be the pulsar period if so.
Put otherwise, what we try to answer here is whether a SED of an unidentified source which is plausibly ascribed to being generated by a pulsar, can be used to determine a range of preferred pulsar periods.
If successful, this technique could allow improving the sensitivity of blind searches of the pulsar period, by determining a limited range of preferred periods to span.
In this work, and as testbed for the concept, we shall consider known and well-measured pulsar spectra blinding us from the knowledge of their timing properties.

In Section 2, we briefly summarize the radiative model already described in detail in our previous studies \cite{compact_formulae,Vigan_2015b,diego_solo,systematic_2019}. In Section 3, we comment on how we adapt it to search for the pulsars' period and
show the results from a sample of 13 pulsars. In Section 4, we draw the main conclusions.

\section{Conceptual introduction of the underlying model}
\label{model}

Our approach must necessarily assume an underlying physical model to describe pulsars' SED.
Here we make use of a synchro-curvature radiative model that has been systematically applied already to fit the known population of high-energy pulsars \citep{Vigan_2015b,diego_solo,systematic_2019}.
Conceptually, the model is based on a self-consistent calculation of the dynamics and radiative properties of charged particles traveling along  magnetic field lines in the outer magnetosphere of a pulsar. 

The timing properties of a particular pulsar, $P$ and $\dot P$, define estimates of the light cylinder radius ($R_{lc} = {c P}/{2 \pi}$), the surface magnetic field
$B_{\star} = 6.4 \times 10^{19} (P[s] \dot{P}[s/s])^{1/2} $ G, and the rotational energy  $\dot{E}_{rot} = 3.9 \times 10^{46} P[s]^{-3} \dot{P}[s/s]$ erg/s.

Along the trajectory, particles are accelerated by an electric field parallel to the magnetic field, $E_{\parallel}$, which is assumed to be uniform in the whole region of particle acceleration, and emit via synchro-curvature radiation losses \citep{cheng_zhang, compact_formulae}. 

At each position of the trajectory, characterized by a given curvature radius of the field lines $r_c$ and local magnetic field $B$, one can compute the Lorentz factor $\Gamma$ and the pitch angle $\alpha$ by solving the particles' equations of motion described in Appendix \ref{app:model}.

In our model, the curvature radius of the magnetic field lines is parametrized as $r_c = R_{lc} (x/R_{lc})^{\eta}$,  where $x$ is the physical distance along the magnetic field line from the star's surface, and $x_{in}$ marks the inner boundary of the region.  The outer boundary is given by $x_{out}$.  We fix $\eta=0.5$, $x_{in}=0.5 R_{lc}$ and $x_{out}=1.5 R_{lc}$. As seen in our previous studies, these three parameters have no relevant effect on the SED on a wide range of reasonable values. The magnetic field is also parametrized as a function of the timing properties and of the magnetic gradient, $b$, a free parameter: $B = 6.4 \times 10^{19} (P[s] \Dot P [s/s])^{1/2} (R_\star/x)^{b}$.

These magnetospheric ($r_c,B$) and kinematic ($\alpha,\Gamma$) magnitudes in turn determine the emission of a single particle at every position, by means of the synchro-curvature radiation formulae for a single-particle SED, $\frac{d P_{sc}}{dE}(\Gamma,\alpha,r_c,B)$, defined in detail in Appendix \ref{app:model}. For details about the typical particles kinematics, see e.g. Section 3 of \cite{compact_formulae} or Section 3 of \cite{Vigan_2015}.
 
The contributions from each trajectory's position to the total, observed spectra are weighted by an effective particle distribution that parametrizes the underlying complexity of the scenario (see further discussion e.g. in \cite{Vigan_2015}). 
In these previous studies, we found that the following parametrized distribution works well for most of the pulsars: 
\begin{ceqn}
    \begin{equation}
        \frac{dN}{dx} = N_0 \frac{e^{-(x-x_{in})/x_0}}{x_0 (1-e^{-(x_{out} - x_{in})/x_0})},
        \label{eq:effective_particle_distribution}
    \end{equation}
\end{ceqn}
where $x_0$ is a length scale: the larger $x_0$ is, the more uniform is the distribution of particles emitting towards us.
The normalization $N_0$ represents the total number of particles directed to us, and does not change the SED shape.

The total emission by synchro-curvature radiation from the whole acceleration region can thus be written in the following simplified way:
\begin{ceqn}
    \begin{equation}
        \frac{dP_{tot}}{dE} = \int_{x_{in}}^{x_{out}} \frac{dP_{sc}}{dE} \frac{dN}{dx} dx.
        \label{eq:convolved_power_spectra}
    \end{equation}
\end{ceqn}   
In a nutshell, the shape of the SED is fully determined by three free parameters (plus the normalization $N_0$): the parallel electric field $E_{\parallel}$, the contrast parameter $x_0$, and the magnetic gradient $b$. $N_0$ is found analytically via a linear regression with the observational spectral data for each set of the other three free parameters.

Thus the amount and type of the radiation emitted is linked to the timing properties, $P$ and $\dot{P}$, which play a crucial role in the spectral determination.

\section{Results}
\label{results}

\subsection{Recognizing the spectral impact of $P$ and $\Dot P$}
\label{results1}

    \begin{figure*}
        \includegraphics[width=0.33\textwidth]{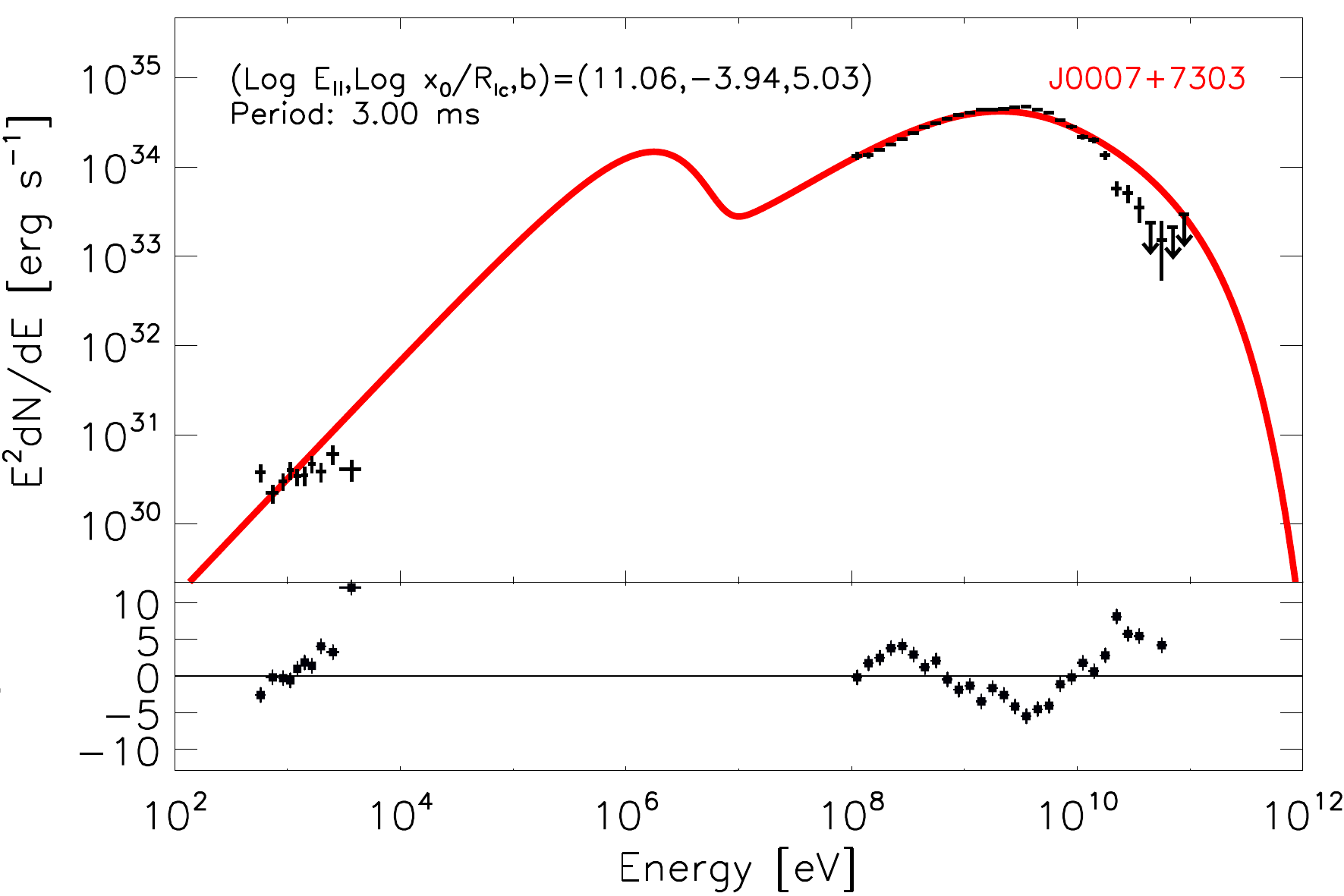}%
        \includegraphics[width=0.33\textwidth]{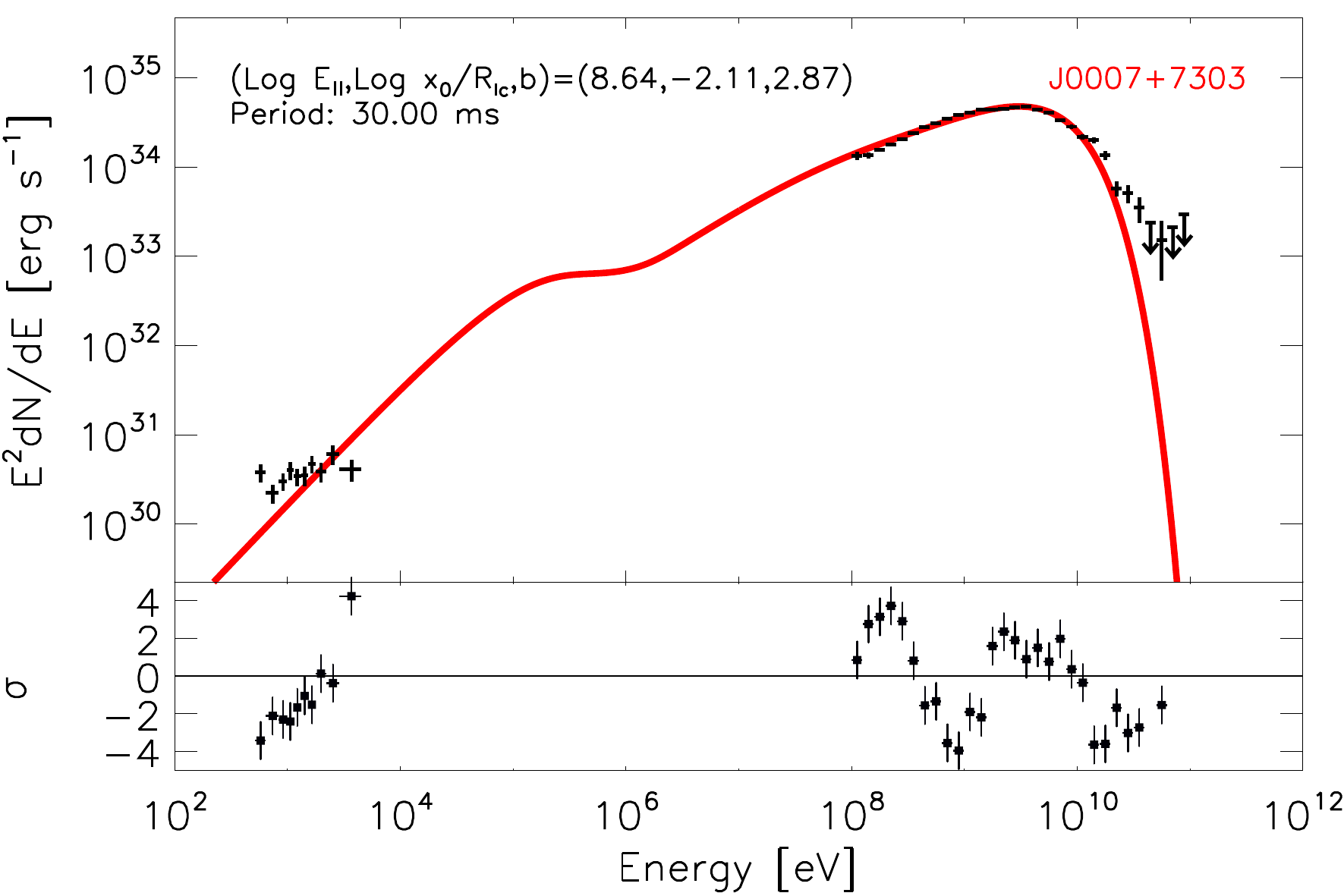}%
        \includegraphics[width=0.33\textwidth]{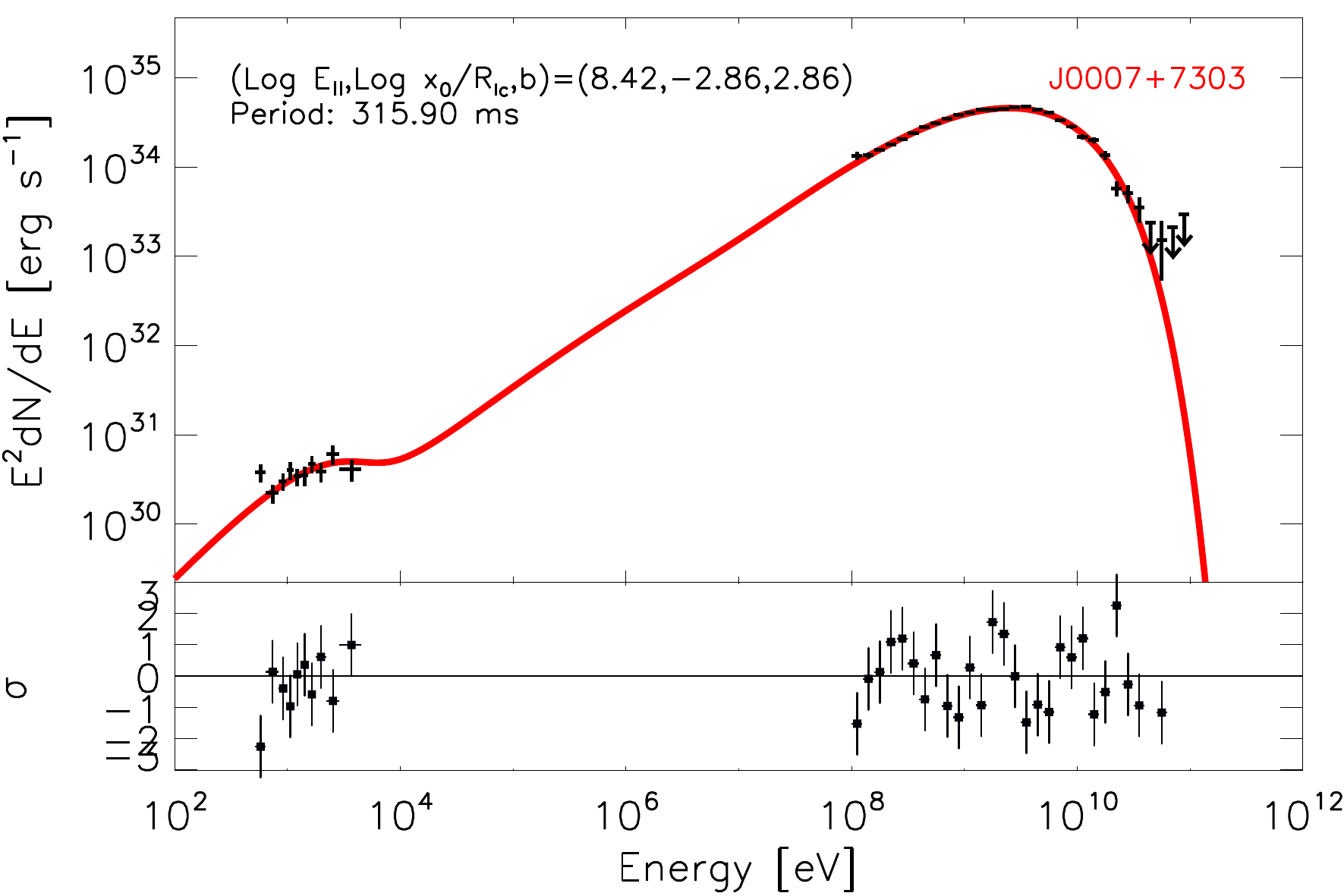}\\
        \includegraphics[width=0.33\textwidth]{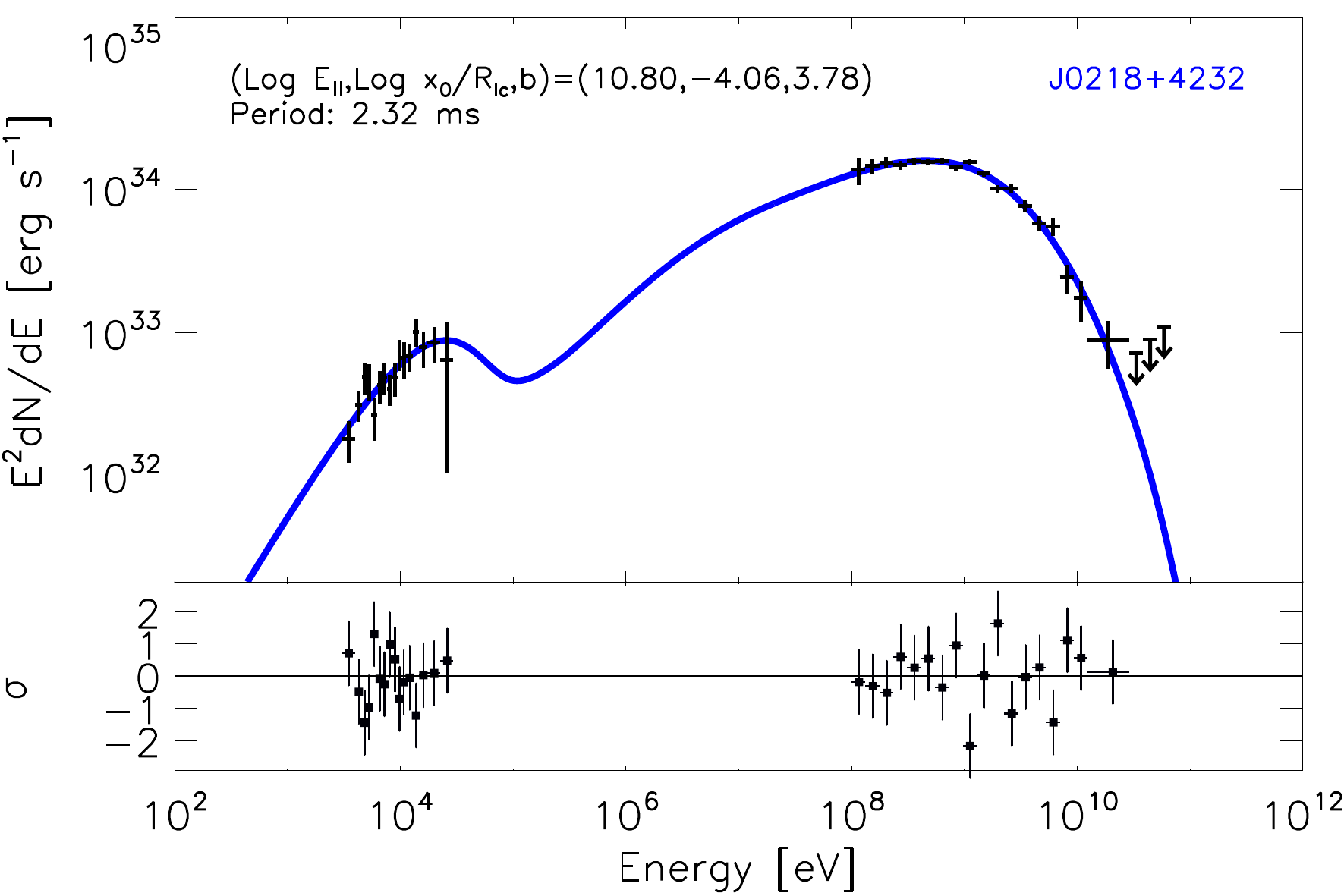}%
        \includegraphics[width=0.33\textwidth]{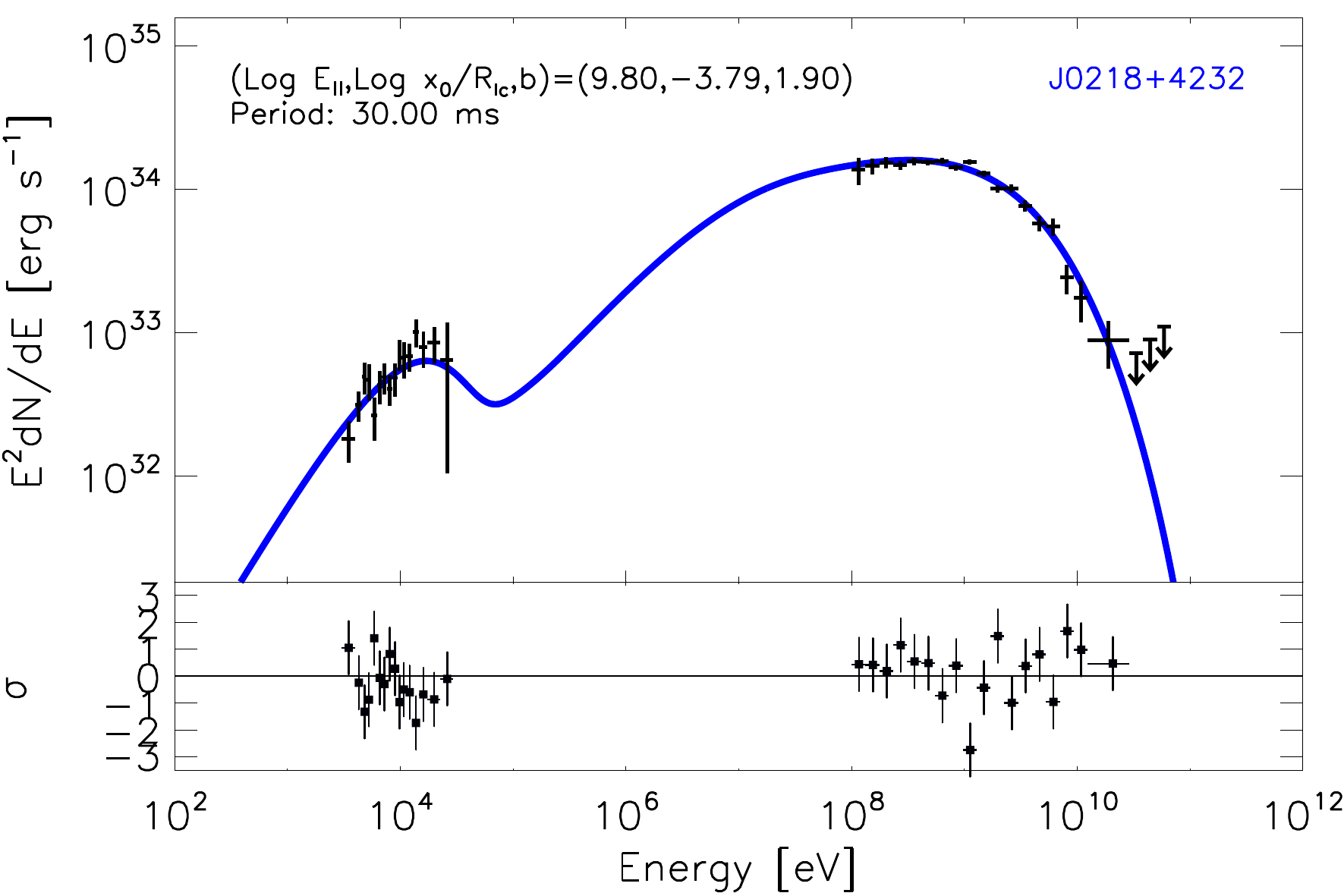}%
        \includegraphics[width=0.33\textwidth]{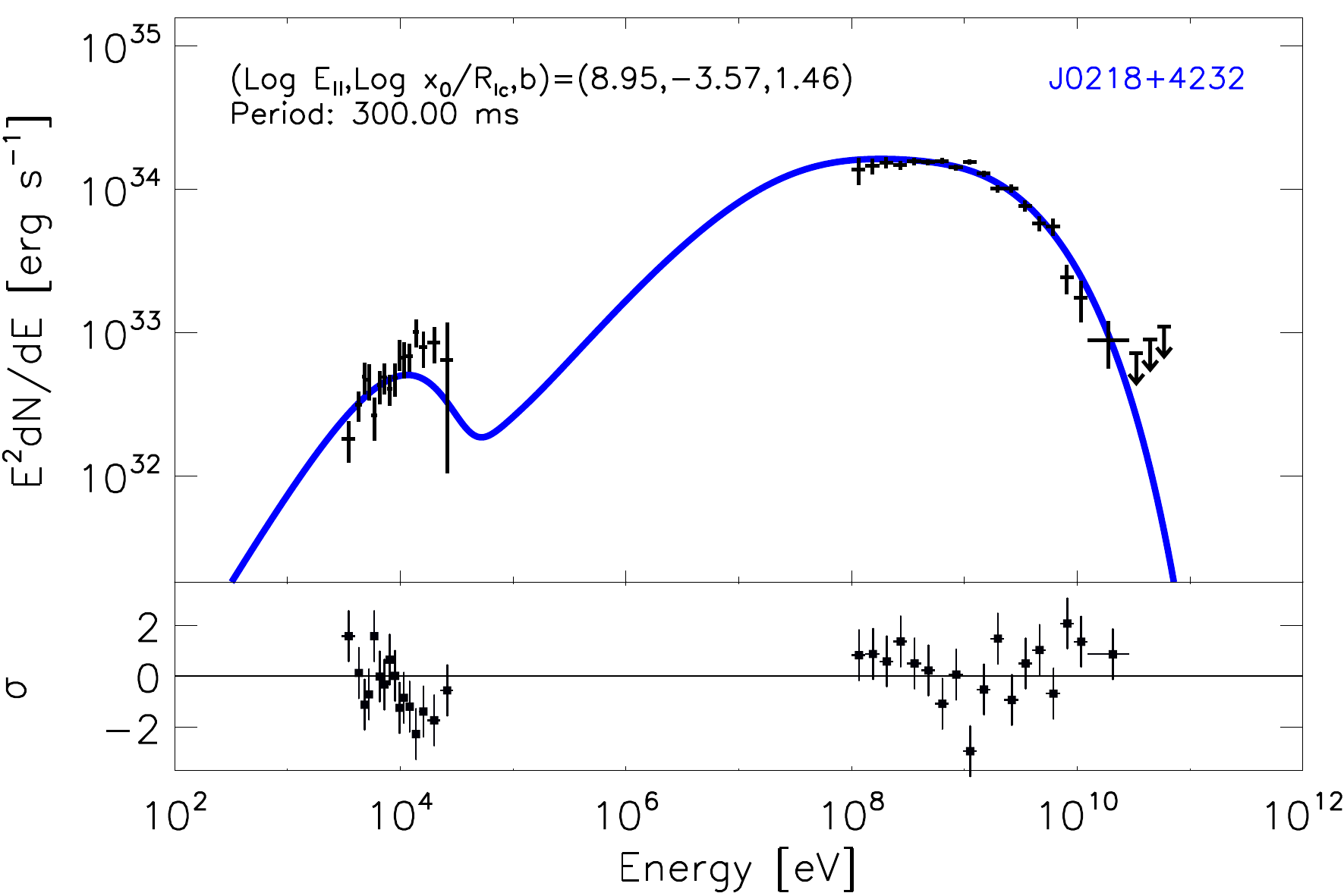}
        \caption{Observed and best-fit model spectra spectra of J0007+7303 (first row) and J0218+4243 (second row). Hereafter, we use red and blue colors for young pulsars and millisecond pulsars, respectively. For each pulsar, we look for the best-fit model assuming three different period, as follow, from left to right: (top) $3$ ms, $30$ ms and the real period (315.9 ms); (bottom): the real period (2.32 ms), $30$ ms and $300$ ms.}
        \label{fig:0007_0218_spectra}
    \end{figure*}

We start by showing how the period $P$ affects the predicted SED of a pulsar.
We consider J0007+7303 assuming three different periods: $3$ ms, $30$ ms, and $315.9$ ms (the real period).
The period derivative is unchanged in this test, and fixed to the measured one for this pulsar.
Assuming each of the values of $P$, we 
draw the best-fit theoretical spectrum produced by the synchro-curvature radiation model 
in Fig. \ref{fig:0007_0218_spectra}.

Note that as in previous works \citep{systematic_2019}, we need a  non-uniform effective particle distribution, $dN/dx$ (small values of $x_0$). 
This means that most of the radiation detected comes from the inner parts of the trajectories that particles run after injection, where the emitting particles still have a relevant perpendicular momentum, and thus their emission is dominated by synchrotron radiation. This is visible as a bump in the X-ray range.

The shape of the three theoretical spectra are different (they have reduced $\chi^2$ values, $\overline{\chi^2}$, equal to 15.01, 6.03, and 1.16  for the $3$ ms, $30$ ms and the real period, respectively).
It can also be seen that the set of parameters producing the best fit (shown in the upper left corner of each panel) is different in each case.

As a second example, let's consider a millisecond pulsar, J0218+4243, in the second row of Fig. \ref{fig:0007_0218_spectra}.
The behavior is similar in general.
In this case, the $P$ considered are: 2.32 ms (the real period), $30$ ms and $300$ ms.
The best fits for each of these periods have $\overline{\chi^2}$ of $0.79$, $1.05$, $1.56$, respectively), and as can be seen in Fig. \ref{fig:0007_0218_spectra}, they look more similar to one another than when we changed the period for the normal pulsar case.  In this case, therefore, the period could not be constrained. This can be ascribed mainly to the shape and quality of the X-ray data: the larger are the errors, the less constraining is the model.

A similar test can be done for the period derivative $\dot P$, now keeping $P$ fixed to its real value.
As an example, we assume now the measured $\dot P$ ($3.57 \times 10^{-13}$ s/s, for J0007+7303), and two values, one order of magnitude above and below the real one, respectively, to test the impact of this change.
The three best-fitting theoretical spectra are almost identical, and would superpose to that shown in the upper right panel of Figure \ref{fig:0007_0218_spectra}.
Hence, a change even across several orders of magnitude of $\dot P$ does not significantly affect the shape of the SED. 

The reason why this happens can be understood by looking at the best-fit parameters for the three $\Dot P$-assumptions in Table \ref{tab:best_fit_parameters_different_pdot}.
There is a degeneracy between $\dot P$ and the magnetic gradient $b$, because both can compensate each other to have the same local
magnetic field in the accelerating region.

Note that, if we only use gamma-ray data, the SED resulting from any change in $P$ and $\dot{P}$ can be fitted by other combinations of the best-fit parameters ($E_{\parallel}, x_0$), while $b$ cannot be constrained.
This is not the case when the X-ray data are included.
Since X-ray spectra heavily depend on the local magnetic field, the parameter $b$ has a strong impact and the degeneracy often breaks, as seen above for J0007+7303 (Fig. \ref{fig:0007_0218_spectra}, top row).

\begin{table}
	\centering
	\caption{Best-fit parameters of J0007+7303 for three different values of $\dot P$.}
	\label{tab:best_fit_parameters_different_pdot}
	\begin{tabular}{cccccc} 
		\hline \hline
		 $\log{\dot P}$ & $\log{E_{\parallel}[\frac{V}{m}]}$ & $\log{\frac{x_0}{R_{lc}}}$ & b & $\overline{\chi^2}$ & Local $B$ [G] \\
		 \hline
		 $-13.45$ & $8.42$ & $-2.87$ & $2.69$ & $1.13$ & $1.27 \cdot 10^5$ \\
		 $-12.45$ & $8.42$ & $-2.86$ & $2.89$ & $1.12$ & $1.29 \cdot 10^5$ \\
		 $-11.45$ &  $8.42$ & $-2.86$ & $3.03$ & $1.12$ & $1.29 \cdot 10^5$ \\
		\hline
		\hline
	\end{tabular}
\end{table}

\subsection{$P$ as a free parameter }
\label{results2}

When both $P$ and $\dot P$ are unknown, they must be searched at once in the framework of this model.
Thus, we have modified the algorithm used to produce the former fittings (that accounts for variations in only three variables  $E_{\parallel}$, $x_0$, $b$, for fixed values of $P$ and $\dot P$, see \cite{diego_solo}) to deal with the period as an additional free parameter.
We consider a logarithmically-spaced period grid from $1$ ms to $1$ s.
For defining $\dot P$ in each instance, we take into account the observed $P-\dot{P}$ diagram.
No $\gamma$-ray pulsars have been found to have small periods (of the order of a few milliseconds) and large period derivatives (between $10^{-15}-10^{-11}$ s/s) or viceversa (periods of $\cal O$(0.1) s, and period derivatives between $10^{-21}$ and $10^{-19}$ s/s). 
Therefore, if we were to change both the period and the period derivative of a pulsar in an unrelated way, we may be imposing unphysical conditions, or at least defining putative pulsars that have not been observed yet.
Instead, we shall simply consider as a proxy
a linear relation between $P$ and $\dot P$ fitting the observed $\gamma$-ray pulsars (using the pulsars' sample presented in \cite{2fpc}):

\begin{equation}
\log_{10}\Dot{P}[s~s^{-1}] = 3.55 \cdot \log_{10}P[s] - 10.5
\label{eq:pdot_formula}
\end{equation}
Particularly, considering the dispersion of the real pulsar data from the linear fit, we are in no way assuming these so-determined $\dot P$-values are always close to the real ones.
\begin{figure}
    \centering
    \includegraphics[width=1.\columnwidth]{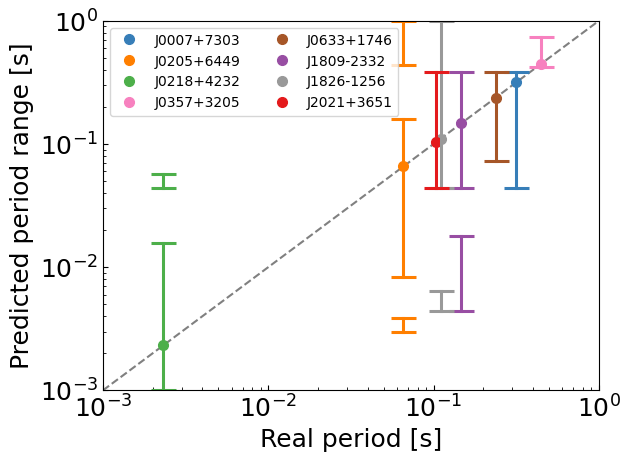}
    \caption{Predicted $P$ ranges for pulsars when 
    such ranges can be defined with our method. When more than one 
    period range is singled out by our model (e.g., as for J0205+6449), we mark it in the plot too.}
    \label{fig:predicted_period_ranges}
\end{figure}
This is an approximation that has as its only aim to define the order of magnitude of $\dot P$ given each of the $P$-values from the grid, as this is thought to be good enough for the results we are after. We have seen in the previous section that the deviation of a real putative $\Dot P$ from the value given by the formula has to be considerably large to significantly modify conclusions.
For each of these periods and their corresponding period derivatives, we perform a spectral fitting 
to the observational data varying the remaining (not-timing) parameters, obtaining a $\overline{\chi^2}$ value.

For defining a testing sample, we have considered the population of high-energy pulsars presented in \cite{CotiZelati20} and selected those pulsars possessing enough data points in both the X-ray and gamma-ray bands to define the spectrum well.
In addition, in order not to complicate the analysis 
we downselected the sample to those pulsars for which a single set of parameters $(E_\parallel, x_0, b)$ is a reasonable description of the data ($\overline \chi^2 \lesssim 1.5)$. 
In this way, we can be confident that our results are not greatly affected by biases related to the bad quality of the data nor that we are using knowledge of $P$ and $\dot P$ to bias the underlying model.

\begin{figure*}
            \centering
            \includegraphics[width=0.33\textwidth]{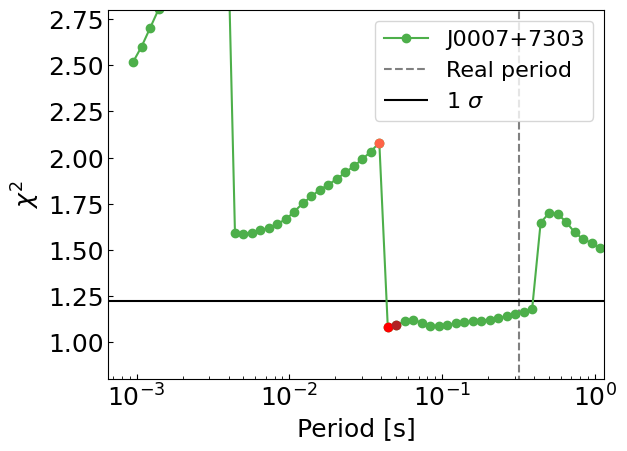}%
            \includegraphics[width=0.33\textwidth]{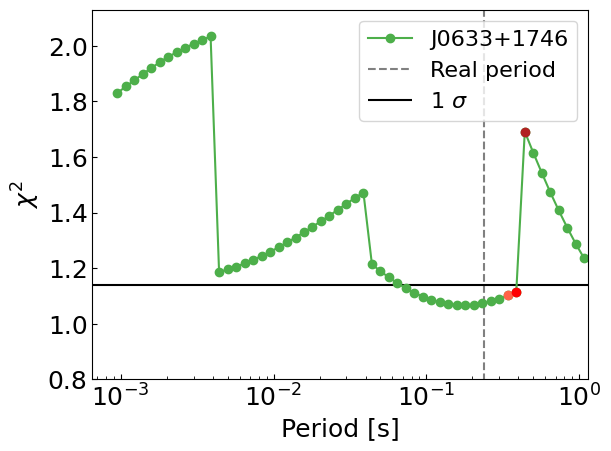}%
            \includegraphics[width=0.33\textwidth]{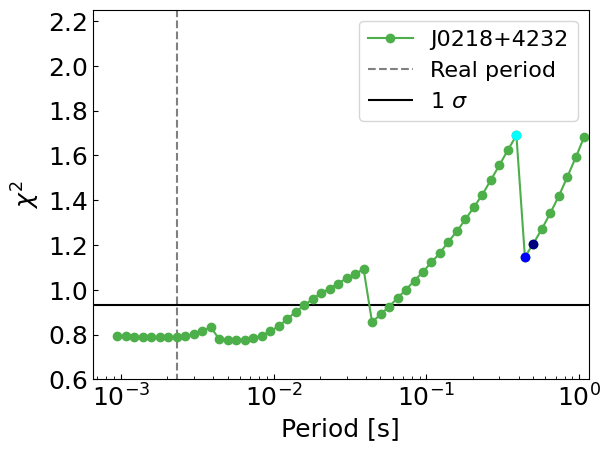} \\
            \includegraphics[width=0.33\textwidth]{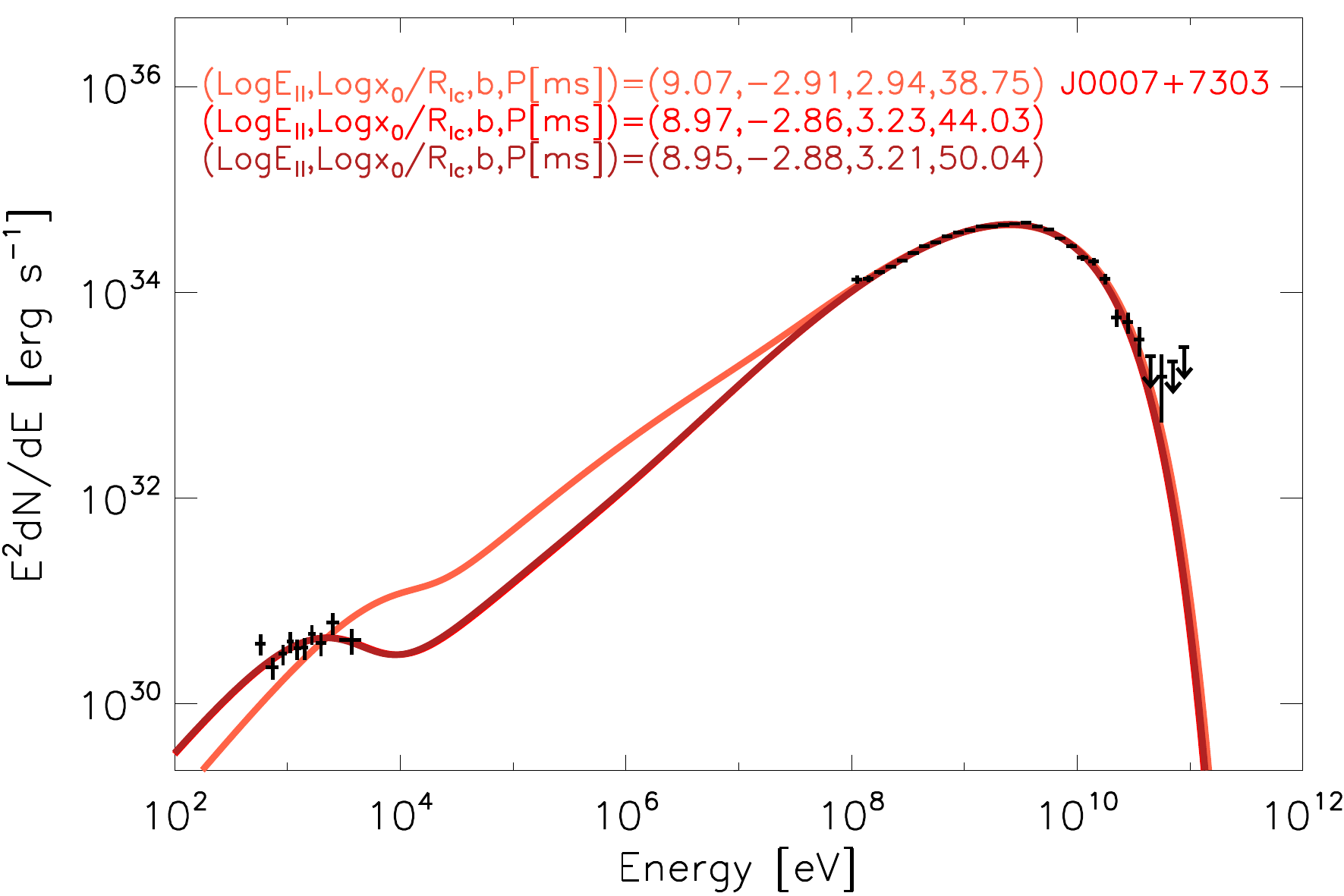}%
            \includegraphics[width=0.33\textwidth]{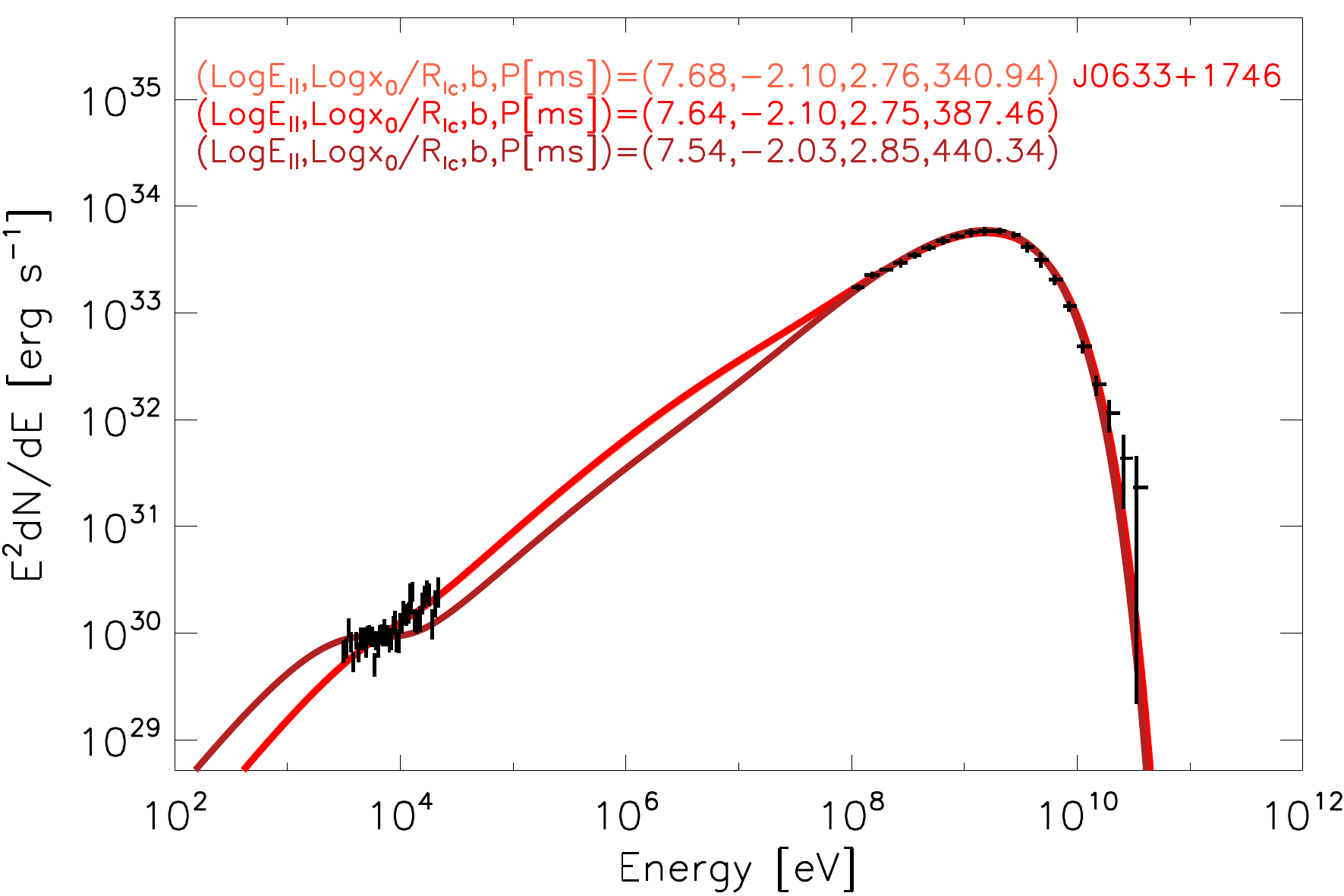}%
            \includegraphics[width=0.33\textwidth]{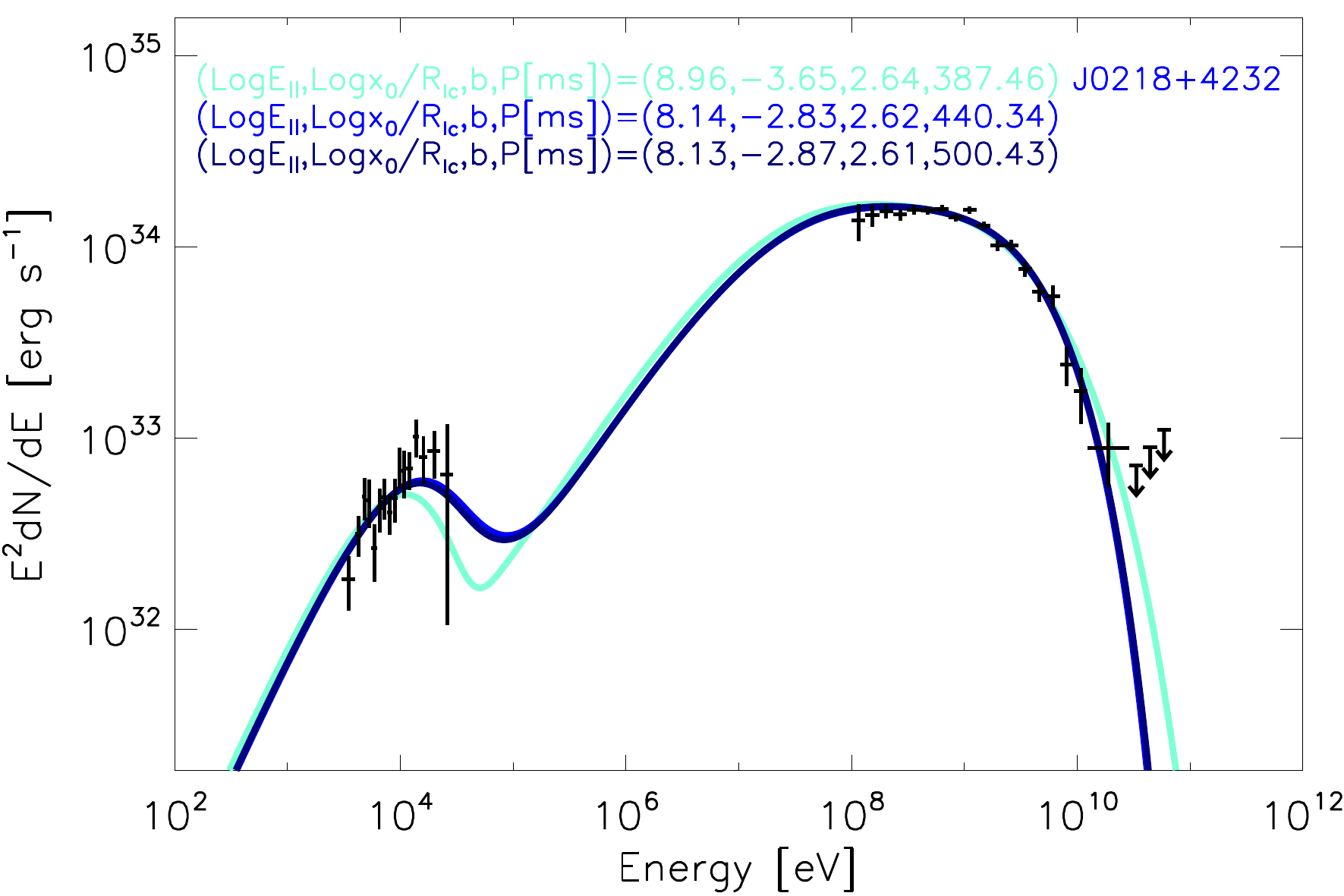} \\
          \caption{Plots of the results of the application of the method for three representative pulsars, from left to right: J0007+7303, J0633+1746 (Geminga) and J0218+4232.
        First row: plots of $\overline{\chi^2}$  versus P. The real period is indicated by dashed gray vertical lines. The horizontal black lines indicate the $1 \sigma$ limit as defined in the text.
        The colored points correspond to the periods around some discontinuities in the plot, for which the corresponding SEDs are plotted in the second row. 
        For the periods on the same side of the discontinuity, the spectra practically overlap.
              }
        \label{fig:chi2_vs_p}
    \end{figure*}

Panels on the top row of Figure \ref{fig:chi2_vs_p} shows plots of 
the $\overline{\chi^2}$ against putative periods $P$ for some examples out of the sample of 13 pulsars selected.
In order to determine a range of statistically plausible periods around the minimum $\overline{\chi^2}$, we follow \cite{avni_factor}.
Such range is identified by finding a threshold $\overline{\chi^2}$, determined by the number of free parameters of the model and the confidence level required (in our case, $4$ parameters and $1 \sigma$ interval), below which the fits are statistically similar.

    \begin{figure*}
            \centering
            \includegraphics[width=0.33\textwidth]{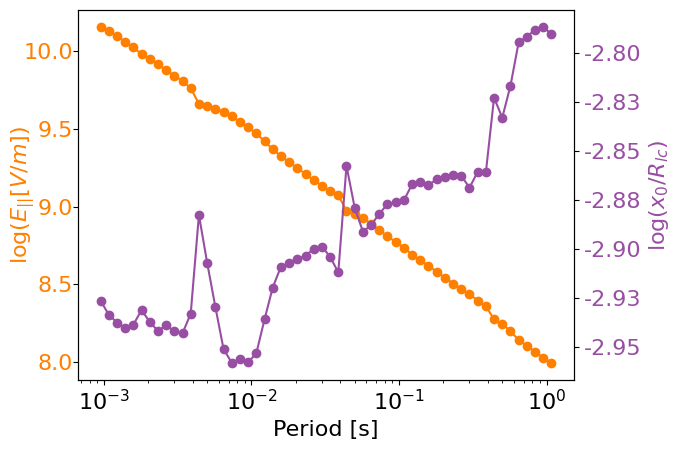}%
            \includegraphics[width=0.33\textwidth]{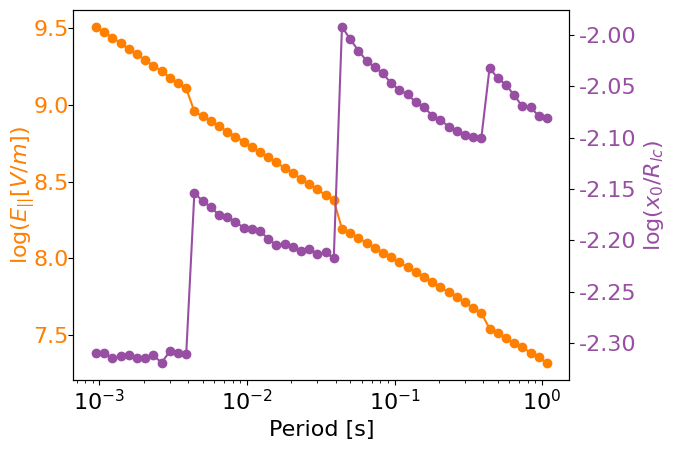}%
            \includegraphics[width=0.33\textwidth]{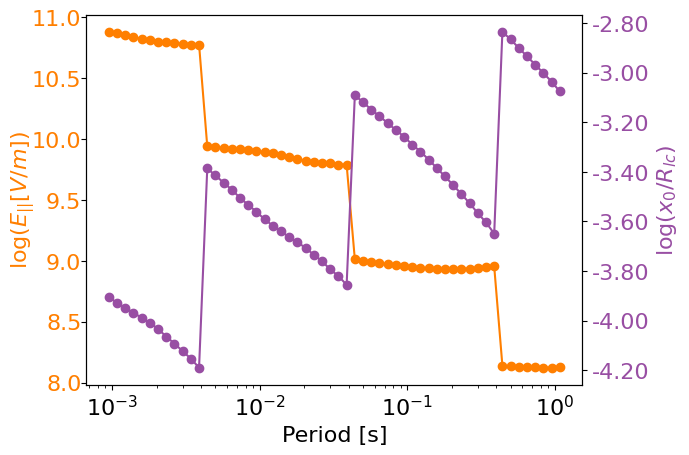}\\
            \includegraphics[width=0.33\textwidth]{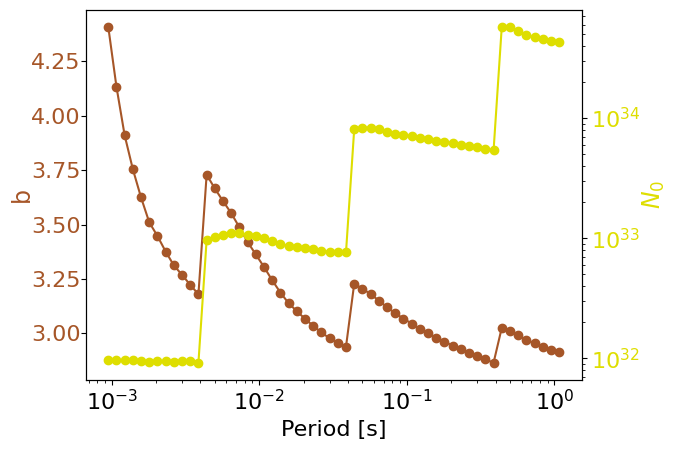}%
            \includegraphics[width=0.33\textwidth]{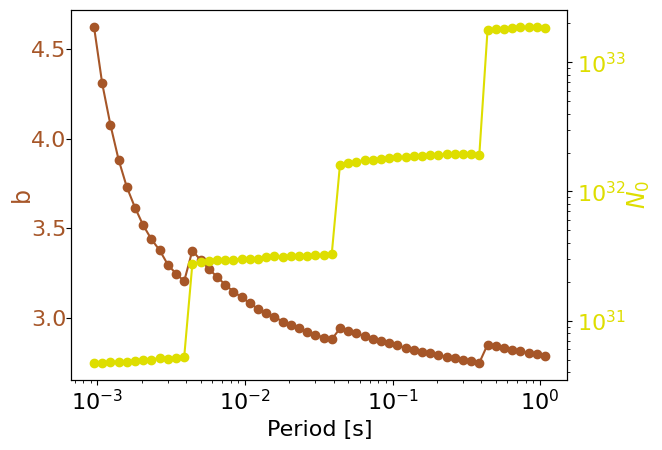}%
            \includegraphics[width=0.33\textwidth]{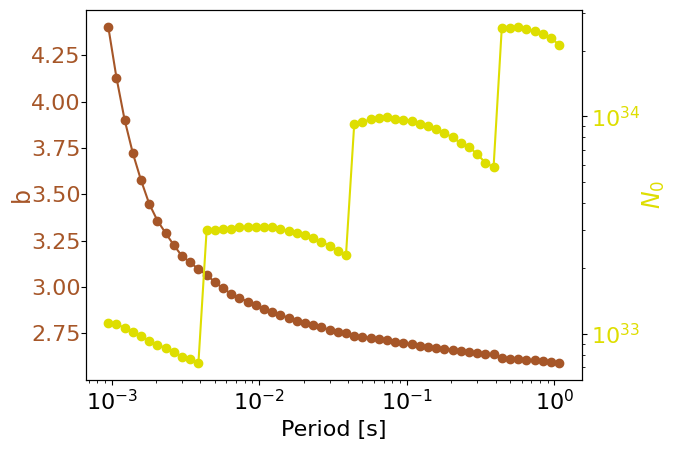}
        \caption{Best-fit parameters as a function of $P$, corresponding to the three pulsars of the top panel of Fig.~\ref{fig:chi2_vs_p}. The panels show joint plots of: $E_{\parallel}$ and $x_0/R_{lc}$ versus $P$ (top); and $b$ and $N_0$ versus $P$ (bottom).
        }
        \label{fig:panels}
    \end{figure*}

For a majority of the pulsars studied
we have found indications 
of the real period as a result of our analysis,
showing relatively small ranges in the plot of $\overline{\chi^2}$  versus $P$ in which the real period is included.
The pulsars in this situation are J0007+7303, J0205+6449, J0218+4232, J0357+3205, J0633+1746 (Geminga), J1809-2332, J1826-1256 and J2021+3651.
Putative pulsar' periods in this range are those for which a good fit to the observational SED can be found, and thus, a priori, one could have concluded that one of them could be the real period if this is unknown.
The width of these regions varies from pulsar to pulsar, but typically span one order of magnitude in $P$, with a couple of cases in which the range is even smaller.
For two of these pulsars (J0218+4232 and J1826-1256) the plot of $\overline{\chi^2}$ versus $P$ exhibit a barrier at a certain period, which can be considered as an upper or a lower limit for the period.
This also defines a range of preferred periods, those between the lower or upper edge of the total period range and the barrier, beyond which a similarly good fit cannot be found.
The size of the so-determined regions is similar as before.
The constrained period ranges for our sample are shown in Fig. \ref{fig:predicted_period_ranges}.

Still, for some of the pulsars selected (J1420-6048, J1513-5908, J1747-2958, J1838-0537 and J2021+4026), the results are non-conclusive, i.e., do not give any information regarding $P$ as
essentially all points in the $\overline{\chi^2}$ versus $P$ plots lie below the limit defined by the threshold $\overline{\chi^2}$.

Comparing the total luminosity radiated for the best-fit obtained at each value of $P$ with the energy budget (given by the spin-down power corresponding to that period), we could in principle obtain additional constraints.
However, this affects only three of the pulsars studied. In these cases the period ranges singled out by our model are further reduced, improving the overall result for them.
The reason that this is not the general case is to be found in the fact that the spin-down power carries with it the uncertainty assigned to $\dot P$ in the search.

The panels in the second row of Figure \ref{fig:chi2_vs_p} show how the best-fitting SED changes for some regions where $\overline{\chi^2}(P)$ is discontinuous (see the colored dots in the first row). 
We proceeded to study in detail the grid around this region and to study in detail the best-fitting SED in each case.
The panels of Figure \ref{fig:panels} show the variation of the best-fit parameters ($E_{\parallel}$, $x_0$ and $b$, and the normalization of the particle distribution $N_0$) as a function of $P$. 
Note that in all the panels of Figure \ref{fig:panels} there are some abrupt changes of the parameter selected by the best fits for each given period (which correspond to the jumps of $\overline{\chi^2}$ in the top row of Figure \ref{fig:chi2_vs_p}).
This is a result of the underlying model trying to accommodate itself to represent in the best possible way an increasingly uncomfortable situation.

Similar trends are observable in all panels of Fig. \ref{fig:panels}.
For instance, the logarithm of the parallel electric field needed to reproduce a given SED linearly decreases with $P$. 
This is consistent with the results of  \citep{Vigan_2015b,diego_solo}, that showed how the millisecond pulsars were found to have higher best-fit $E_{\parallel}$ than the standard pulsars ($P>10$ ms).
There is a physical reason for this to happen.
A smaller $P$ implies a smaller light cylinder radius, and a larger curvature radius $r_c$.
This implies stronger radiative losses, which are counteracted by a large parallel electric field on the region, induced by the fast rotation.
For large $P$ the situation is the contrary: larger light cylinder radius and thus relatively small $r_c$, which provokes not so strong radiative losses, thus a large $E_{\parallel}$ is not needed.
However, the latter can only work up to extent that the accelerating region is powerful enough to make particles able to emit in $\gamma$-rays. For sufficiently large periods, this is no longer possible and the preferred period range is limited.

The same panels show the variation of $x_0$. 
Unlike $E_{\parallel}$, $x_0$ slightly increases.
However, if we plot the physical value of $x_0$ (not normalized to $R_{lc}$), using the corresponding light cylinder for each pulsar period in the grid, a clear anticorrelation with $E_{\parallel}$ can be seen. 
This was observed (see \cite{Vigan_2015b,diego_solo}) when comparing different pulsars. Varying the period within a single pulsar, while conserving the observed SED, surmises this same effect.
The best-fit values of $x_0$ are small, indicating that most of the radiation emitted comes from the beginning of the particles' trajectories in all cases.

Finally, the panels in Figure \ref{fig:panels} also show how the magnetic gradient $b$ and the normalization $N_0$ vary with $P$.
$b$ and $P$ define the local strength of the magnetic field along the acceleration region.
In fact, this would depend also on $\dot P$ (see section 2), but recall that 
we are herein adopting a dependence of the latter with $P$, as described.
Thus, the parameter $b$ decreasing with an increasing $P$ allows to maintain the same local magnetic field in the accelerating region.
On the contrary, $N_0$ increases with $P$.
Again, if $P$ is small, the light cylinder is close to the neutron star and thus the curvature radius is large.
This implies substantial radiative losses for each traveling particle.
Therefore, a relatively small amount of particles is needed to emit the detected flux, and the value of $N_0$ is relatively smaller than when a higher $P$ is adopted.
%

\section{Discussion and conclusions}
\label{conclusions}

We have studied the impact of the timing parameters of pulsars in a synchro-curvature model
that was earlier shown to deal with the spectral data of all high-energy pulsars. 
Considering the observational high-energy data fixed, we have uncovered how degeneracies arise when trying to fit them assuming timing parameters different 
from the real ones. 
In particular, we have found a degeneracy 
between $P$ and the pair ($E_\parallel,x_0)$ --at a fixed $\dot P$--, and between $\dot P$ and $b$ --at a fixed $P$, that leads to similar SEDs. 
The origin of such degeneracies is related to the physics of the problem. 
For instance, a lower magnetic field in the light cylinder can be obtained by enlarging the period (thus moving the light cylinder away from the star) or reducing the magnetic gradient with which the surface magnetic field is reduced when moving away from it.
When dealing with known pulsars, this does not represent an issue, since the real timing values break these degeneracies. 
However, it may affect the correct determination of the timing parameters of unknown pulsars via a direct fit to the SED, at least in some cases. 
Because of these degeneracies, our analysis shows that if we were to blind ourselves from the knowledge of their periods, the $\gamma$-ray data alone is not able to determine what kind of pulsar is behind it in most cases.
This emphasizes how similar millisecond and standard pulsars are in regards to their $\gamma$-ray emission.

Blinding ourselves from the knowledge of timing parameters of several pulsars, the methodology presented is successful in determining a preferred  period range that includes the real periods in a majority (8/13) of the cases.
The preferred period range is limited in these cases to about one order of magnitude or better.
This encompasses a plausible improvement for blind search algorithms, having to span a smaller range.
For the rest of the pulsars analyzed (5/13), the fits show a degeneracy in the period, and no range could be identified.
This is a direct result of two factors: the shape and/or quality of the data and the underlying degeneracies 
between $P, \dot P$ and the parameters of the model $(E_{\parallel},x_0,b)$.

A concern for the general model application rest on the required usage of X-ray data. 
The method assumes that apart {\it Fermi}-LAT data, we have a possible counterpart in X-rays. 
This is useful to break the degeneracy between $\dot P$ and $b$, given that the latter parameter is sensibly affecting the X-ray regime.
However, differently from $\gamma$-ray data, current pulsar X-ray data do not usually come from a survey, thus we require X-ray pointings towards unidentified sources to exist so that possible counterparts are identified and tested individually.
This is in fact not uncommon for the unassociated Galactic sources existing in the \emph{Fermi}-LAT catalog today.
Moreover, with the advent of eXTP \citep{extp} and Athena \citep{athena} observatories, more homogeneous surveys could be used to identify and test possible X-ray counterparts.

Finally, with the upcoming Third \emph{Fermi} Pulsar Catalog and their improved $\gamma$-ray pulsar data, it will be interesting to apply this methodology to those sources qualified as plausible pulsars, but for which a period is unknown.

\section*{Acknowledgments}
This work has been supported by the grants
PID2021-124581OB-I00, PGC2018-095512-B-I00 as well as 
the Spanish program Unidad de Excelencia ``María de
Maeztu'' CEX2020-001058-M. 
DIP has been supported by CSIC program JAE-ICU.
DFT also acknowledges USTC and  the Chinese Academy of Sciences Presidential Fellowship Initiative 2021VMA0001. 
DV is funded by the European Research Council (ERC) under the European Union’s Horizon 2020 research and innovation programme (ERC Starting Grant IMAGINE, No. 948582).

\bibliography{period_pulsar_sc_spectra}{}
\bibliographystyle{aasjournal}

\appendix
\section{Further details of the synchro-curvature radiation formalism}\label{app:model}

Here we summarize the details of the underlying synchro-curvature radiation model, presented already in \cite{compact_formulae,Vigan_2015,Vigan_2015b,diego_solo}.
We numerically solve the equations of motions of the traveling charged particles:
\begin{equation}
 \frac{d\vec{p}}{dt} = ZeE_\parallel \hat{b} - \frac{P_{sc}}{v}~\hat{p}~,
\end{equation}
where the relativistic momentum $\vec{p}$ has a parallel ($p_\parallel=p\cos\alpha$) and a perpendicular component ($p_\perp = p\sin\alpha$) respect to the magnetic field lines (directed along $\hat{b}$); $P_{sc}=\int (\frac{d P_{sc}}{dE}) dE$ is the synchro-curvature power (where $\frac{d P_{sc}}{dE}$ is defined below), $v\sim c$ the particle velocity and $Ze$ its electric charge. Initial values of the Lorentz factor $\Gamma$ and the pitch angle $\alpha$ are set to typical initial values, $10^3$ and $45^\circ$, respectively (their precise value has negligible effects on the final SED within the range of reasonably expected values).

Solving these equations allows to compute the synchro-curvature radiation emitted by a single particle at a given position, \begin{ceqn}
\begin{equation}
    \frac{d P_{sc}}{dE} = \frac{\sqrt{3}(Ze)^2\Gamma y}{4 \pi \hbar r_{eff}} \left[(1+z)F(y) - (1-z)K_{2/3}(y)\right]
    \label{eq:sc_energy_spectra}
\end{equation}
\end{ceqn}
where:

\begin{ceqn}
    \begin{align}
        F(y) &= \int^{\infty}_y K_{5/3}(y')dy' \\
        r_{gyr} &= \frac{m c^2 \Gamma \sin{\alpha}}{e B} \label{eq:f_y}\\
        \xi &= \frac{r_c}{r_{gyr}} \frac{\sin^2{\alpha}}{\cos^2{\alpha}} \label{eq:xi_def} \\\
        r_{eff} &= \frac{r_c}{\cos^2{\alpha}} \left(1 + \xi + \frac{r_{gyr}}{r_c} \right)^{-1} \label{eq:effective_radius} \\
        Q_2^2 &= \frac{\cos^4{\alpha}}{r_c^2} \left[ 1 + 3\xi + \xi^2 +\frac{r_{gyr}}{r_c} \right] \label{eq:q2}\\
        E_c &= \frac{3}{2} \hbar c Q_2 \Gamma^3 \label{eq:e_c} \\
        z &= (Q_2r_{eff})^{-2} \label{eq:z}
    \end{align}
\end{ceqn}
where: $y$ is the ratio $E/E_c$, with $E$ and $E_c$ the photon energy and the characteristic energy of the emitted radiation, respectively;
$K_n$ are the modified Bessel functions of the second kind of index $n$, the solutions of the Bessel equation with complex argument;
$r_{gyr}$ is the Larmor radius, $e$ and $m$ are the charge and rest mass of the particle and $c$ is the speed of light;
$\xi$ is the synchro-curvature parameter, which indicates whether the emission is dominated by synchrotron or by curvature radiation, or if it is a mixture of both (details in \cite{compact_formulae}).

\end{document}